# Statistical Material Law Support for Room Temperature Superconductivity in the Lead Apatite System

Ning Chen[1*], Yang Liu[1], and Yang Li[2*]
1 School of Materials Science and Engineering, University of Science and Technology Beijing, Beijing, 100083, P.R. China
2 Department of Engineering Science and Materials, University of Puerto Rico, Mayaguez, Puerto Rico 00681-9000, USA
*Corresponding author：nchen@sina.com; yang.li@upr.edu

**Abstract**
Quantum mechanics calculations of an electronic orbital coupling distribution of the new possible Room Temperature (RT) system of lead apatite $Pb_{10}(PO_4)_6O$ system show that a complex orbital coupling characteristics of the new system are similar to those of copper oxide, and with multiple orbital interactions. The overall band width is significantly larger than the current largest copper oxide system, which can also be compared to near RT high-pressure hydride systems. By applying a material dependence law relating overall band width to superconducting critical temperature across various typical superconducting systems, we predict that the electronic structure of the new material can support achieving RT superconductivity in condition of a flatter band.
**Key words**: superconductivity, room temperature superconductor, lead apatite, material statistic law

  Presently, many calculations[1,2,3] of density functional theory have shown the Fermi level of Cu doped lead apatite (or LK-99) systems[4,5] to be a flat band, which is similar to high-temperature superconductivity characteristics and supports the occurrence of high temperature superconductivity. However, the flat band feature alone cannot quantitatively predict the critical temperature of a system. Additionally, although the mechanism of high-temperature superconductivity remains controversial, the possibility of room temperature (RT) superconductivity is questioned by almost all superconductivity mechanisms. Regarding experiments, there are many influencing factors in the synthesis process of partially Cu substituted $Pb_{10}(PO_4)_6O$, resulting in differences in samples and property performance. Thus, the existence of superconductivity cannot be determined conclusively so far. Therefore, it is important to seek new evidence for RT superconductivity from an independent method rather than just DFT theories and property measurement experiments.

  The study of big data in materials relies on large amounts of experimental data, but derived laws can be applicable for predicting most new unknown systems. Statistical laws do not rely on any theoretical mechanisms; evenly they can verify whether any theoretical mechanism is correct. Unlike studies on crystal structure, composition, or other macroscopic parameters, our big data work[6] mainly focuses on the characteristics of electronic orbital interactions of superconducting systems, which requires quantum mechanical calculations or approximation method to obtain band structure information of the material first. By comparing and summarizing the electronic structures of typical superconducting systems over the full energy range, we have obtained a relationship between the overall band width parameter and the maximum critical temperature of a superconducting system[7]. Using this statistical law, we can predict the critical temperature of any new system. This work is hoped to answer some basic experimental and theoretical questions on RT superconductivity through the statistical prediction for this new possible RT superconductor.

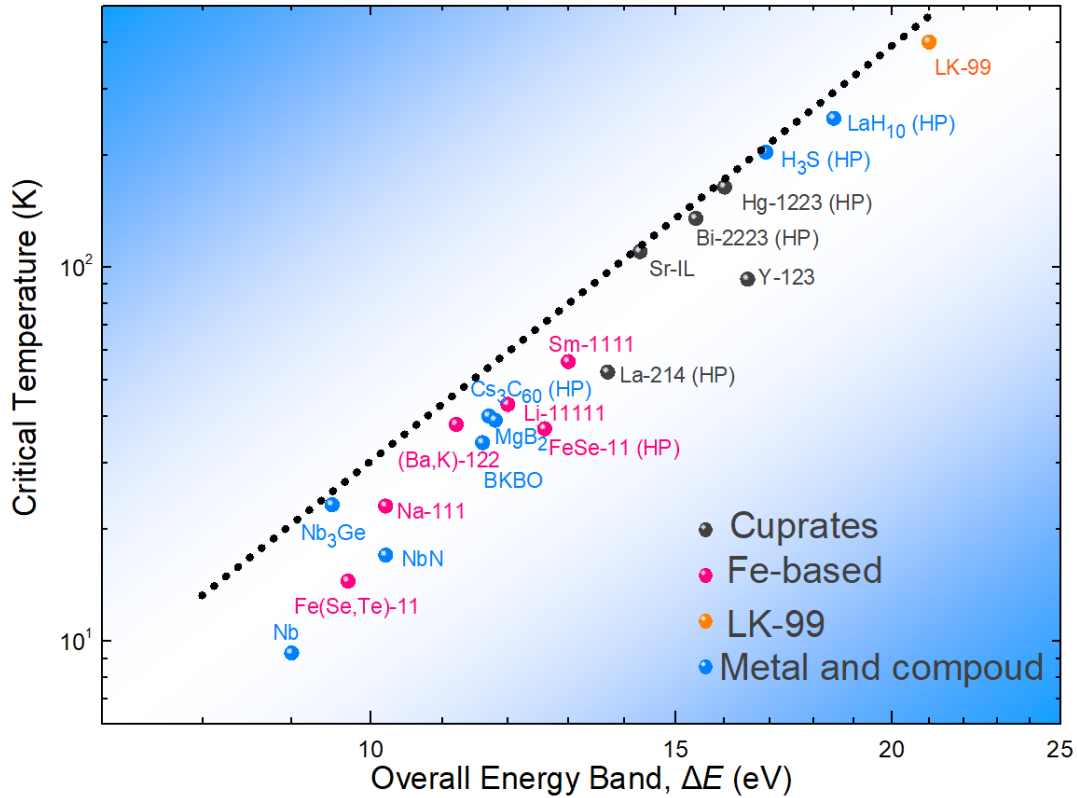

Figure 1 shows the earlier summarized statistical rules, including overall band width on the horizontal axis and critical temperatures (suitable for metals or compounds) or maximum critical temperatures (suitable for iron-based compounds and copper oxides) on the vertical axis.

Although over hundreds of typical superconductors with relatively high critical temperatures have been discovered to date, they can be simply classified into four families with similar energy band structure characteristics and similar superconducting temperatures: pure metals ($T_c<10K$) and intermetallic compounds ($T_c<40K$, also including high-pressure hydrides, $T_c>100K$), as well as high-temperature superconducting systems (iron-based compounds, $T_c=21-55K$, and copper oxides, $T_c=32-132K$). Each system has around a dozen typical examples suitable for analysis, accounting for 70% of all data. By analyzing the energy band characteristics related to orbital interactions, an overall band width can be obtained by ignoring isolated orbital bands, not just the width of the valence or conduction band. Regarding material knowledge, the vicinity of the Fermi level mainly affects the concentration of superconducting carriers, while the overall band structure features may affect the strength of electron pairing. Our statistical law indicates a positive correlation between the overall energy band width for all interacting orbital energy range and the maximum critical temperature. And below the trend line, it reflects the flatness of the energy band. As a statistical law without relying on any superconducting mechanism, it provides an independent statistical analysis conclusion on RT superconductivity problems.

The material dependency law comes from the overall band width of density-of-states curves from electronic structure calculations by density functional theory with the first principles method. All the electronic structure directly reflects interactions between atomic orbitals, coupling, or entangling. To illustrate their physical meaning regarding the trend of critical temperature versus overall band width, here we provide a schematic qualitative description of important interacting atomic orbitals for different kinds of superconductors, as shown in Figure 2:

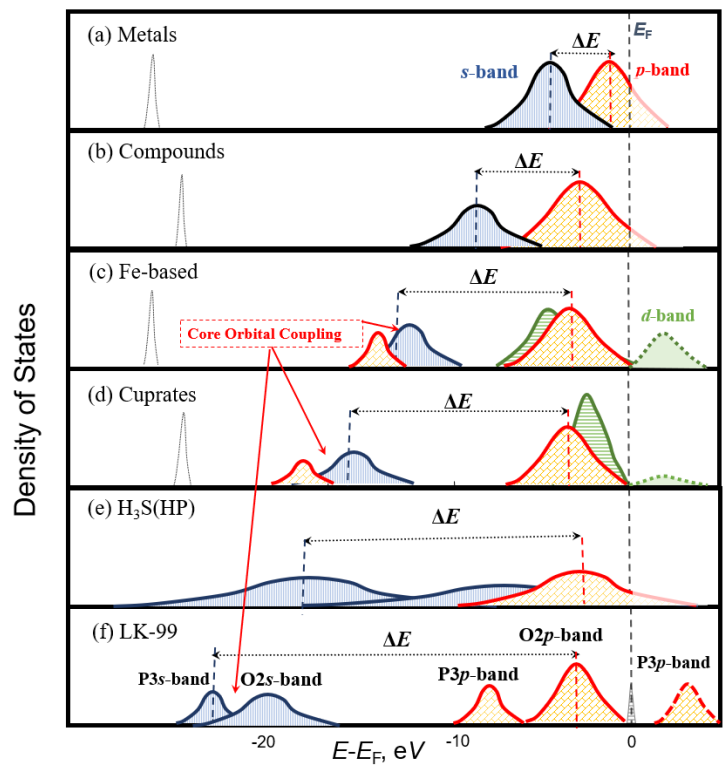

Figure 2 Comparison of overall band width ($\Delta E$) or range of all coupling according to 6 schematic DOS curves corresponding to four typical superconductor classes (Fermi level at 0). (a) and (b) are typical DOS curves for classical compound superconductors and metal superconductors. For both, $sp$ coupling involves only valence orbitals near the Fermi level. (c) and (d) are typical DOS curves for cuprates and iron-pnictides. For cuprates, independent core orbital couplings are around -20eV binding energy, from two core orbitals: the filled $2s$-shell of the oxygen ion on the $CuO_2$ layer and the filled $p$-shell of the interlayer ion neighboring the $CuO_2$ layer. For pnictides, the core orbital coupling corresponds to the As $4s$ core orbital and neighboring Ba or rare earth $5p$ core orbitals (all around -12eV binding energy). (e) and (f) High pressure hydrides and new possible RT system

As shown in Fig. 2a and 2b, metal superconductors usually occupy a smaller energy band width because these metal superconductors have a narrow energy band structure involving one of the following two mixing orbital types: the *s+d* type like Nb and the *s+p* type like Pb, which have been illustrated in the periodic table showing the distribution of chemical elements for which superconductivity has been observed without applied pressure[8].

As shown in Fig. 2c-2d: In contrast to conventional superconductors, there are additional core orbital couplings in the tight binding energy for cuprate systems[9,10] (as shown in Fig. 2d) and pnictide systems[11] (as shown in Fig. 2c), involving *sp* coupling from some core orbitals rather than valence ones. The overall energy band in HTS was extended to -20eV binding energy because of this special sp orbital coupling. Furthermore, as the O2*s* orbital is much deeper than As4s in binding energy, cuprates usually have a larger energy band width than pnictides.

Interestingly, our DFT calculation shows a similar bandwidth of MgB$_2$ (Tc=39K) corresponding to an effective bandwidth of La-214 cuprate (*T*c=40K), and high pressure H$_3$S (*T*c~203K) corresponds to Bi-2223 cuprate (~134K under high pressure). Although MgB$_2$ and H$_3$S, without core orbital coupling, are thought to be conventional superconducting systems, this overall bandwidth law seems to work well for all kinds of superconductors with a flat band feature. Although there are clear differences in the band structure characteristics of typical superconductor systems in different categories, an overall band width (or energy span) has a good positive correlation with the corresponding superconducting critical temperature. With this statistical law, we can predict the superconducting critical temperature of a new system.

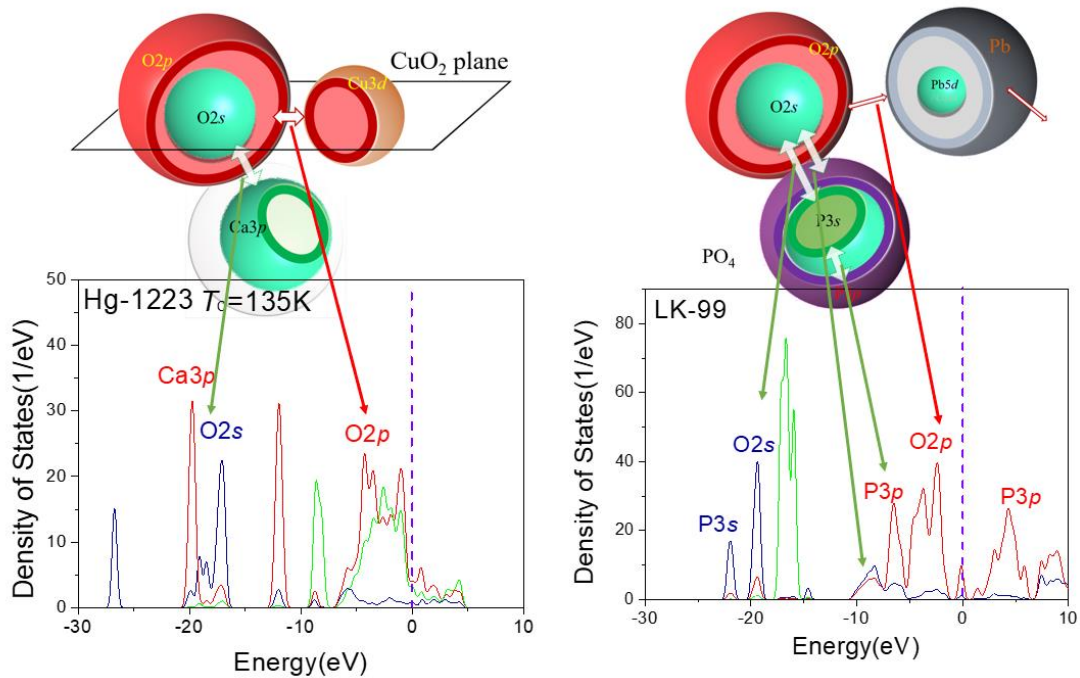

Figure 3 Density-of-States of cuprate and lead apatite, obtained by energy band calculation method.

To compare with the electronic structure characteristics of previous typical systems, the energy band calculation method for the new system was still performed using the CASTEP[12] method in Materials Studio, with the same calculation parameters[13] as previous work. As shown in figure 3, quantum mechanical calculations indicate the overall band structure of the new lead-apatite system also has very interesting characteristics. Similar to copper oxide, it also has a wide inner coupling band, which has a more complex orbital interactions than copper oxide. The DOS of P shows signs of *sp* hybridization, and as a cationic characteristic, the P3*s* orbital is significantly lower than the 2*s* of oxygen ions. If more electrons are introduced, the Fermi surface of the system will rise into the P3*p* band. Even without considering changes in Fermi level position, the new system will have a much wider characteristic band width than copper oxide. Therefore, if we speculate based on the statistical law relating critical temperature and electronic structure characteristics, the critical temperature limit of the new system must exceed that of the highest copper oxide system (138K at atmospheric pressure and 160K at high pressure), suggesting the possibility of RT superconductivity in condition of a flatter band.

Additionally, we also analyzed the details of the orbital coupling in all these typical superconducting systems to discuss other property characters, especially a critical current. According to our summary of energy band characteristics, since traditional metal and compound systems usually have a simple direct *sp* orbital interactions and also the core *s*-peak is larger than *p*-peak, the superconducting current density seems much richer. However, iron-based and copper-based high-temperature superconducting systems, due to their two-layered energy bands with an indirect relation between inner and outer layers, and also *s*-peak intensity is smaller than *p*-peak one, which may cause a lower concentration of superconducting carrier. Furthermore, as a more complex orbital interaction or more inefficient correlation, this feather may cause more weakening sample critical current in LK-99, and the instability and ambiguity of experimental research phenomena might be explained. Therefore, seeking more efficient inner orbital coupling in RT superconductivity to get a better property should thus be our next goal.

In summary, according to our statistical law relating electronic structure and superconducting critical temperature, an independent conclusion has been obtained that the new lead apatite system can support RT critical temperature limit of superconductivity, in condition of a proper flat band. This result is independent of experiment and theory, even superconducting mechanism. We believe the laws of materials are the foundation of the theory of superconductivity, opening a window for us to see the mysterious world of RT superconductivity more clearly. From this perspective, clarifying this law is as important as experimentally confirming a RT superconductor.